\documentclass[11pt]{article}

\usepackage[T1]{fontenc}
\usepackage[utf8]{inputenc}
\usepackage[margin=1in]{geometry}
\usepackage{microtype}
\usepackage{graphicx}
\usepackage{booktabs}
\usepackage{array}
\usepackage{tabularx}
\usepackage{longtable}
\usepackage{multirow}
\usepackage{amsmath,amssymb}
\usepackage{caption}
\usepackage{subcaption}
\usepackage{enumitem}
\usepackage{float}
\usepackage{xcolor}
\usepackage[numbers,sort&compress]{natbib}
\usepackage{hyperref}

\definecolor{Accent}{HTML}{1F4E79}
\definecolor{SoftGray}{HTML}{6B7280}

\hypersetup{
  colorlinks=true,
  linkcolor=Accent,
  citecolor=Accent,
  urlcolor=Accent,
  pdftitle={Orchestrating Human-AI Software Delivery},
  pdfauthor={Maximiliano Armesto and Christophe Kolb}
}

\newcolumntype{Y}{>{\raggedright\arraybackslash}X}
\newcolumntype{R}{>{\raggedleft\arraybackslash}X}

\title{\vspace*{-1.2em}\Large\bfseries Orchestrating Human-AI Software Delivery:\\
A Retrospective Longitudinal Field Study of Three Software Modernization Programs}
\author{Maximiliano Armesto \and Christophe Kolb\\[0.35em]
\small Taller Technologies\\
\small \texttt{maximiliano.armesto@tallertechnologies.com} \quad \texttt{christophe.kolb@tallertechnologies.com}}
\date{March 17, 2026}

\begin{document}
\maketitle
\vspace{-1.2em}

\begin{abstract}
Evidence on AI in software engineering still leans heavily toward individual task completion and repository-level issue solving, while evidence on \emph{team-level delivery} remains comparatively scarce. We report a retrospective longitudinal field study of an industrial platform, Chiron, that coordinates humans and AI agents across four measured delivery stages: analysis, planning, implementation, and validation. The study covers three real software modernization programs -- a COBOL banking migration (approximately 30k LOC), a large accounting modernization (approximately 400k LOC), and a .NET/Angular mortgage modernization (approximately 30k LOC) -- observed across five delivery configurations: a traditional baseline and four successive platform versions (V1--V4).

The benchmark was assembled retrospectively from engineering records and practitioner recall. We therefore separate \emph{observed} outcomes (stage durations, task volumes, validation-stage issues, and first-release coverage) from \emph{modeled} outcomes (raw person-days and senior-equivalent effort under explicit staffing scenarios). Under the benchmark's baseline staffing assumptions, portfolio totals move from 36.0 to 9.3 summed project-weeks between the traditional baseline and V4; modeled raw effort falls from 1080.0 to 232.5 person-days; modeled senior-equivalent effort falls from 1080.0 to 139.5 SEE-days; validation-stage issue load falls from 8.03 to 2.09 issues per 100 recorded tasks; and task-weighted first-release coverage rises from 77.0\% to 90.5\%.

The longitudinal trajectory is informative. V1 and V2 compress analysis and planning before downstream quality recovers. V3 and V4 add acceptance-criteria validation, repository-native review, and hybrid human-agent execution, and those later configurations simultaneously improve delivery speed, coverage, and downstream issue load. Because the study is observational, retrospective, and confined to a single organization, the results should be interpreted descriptively rather than causally. Even under that conservative reading, the evidence is consistent with a central thesis: the largest gains from agentic software delivery appear when AI is embedded in an orchestrated workflow rather than deployed only as an isolated coding assistant.
\end{abstract}

\begin{figure}[H]
\centering
\includegraphics[width=\textwidth]{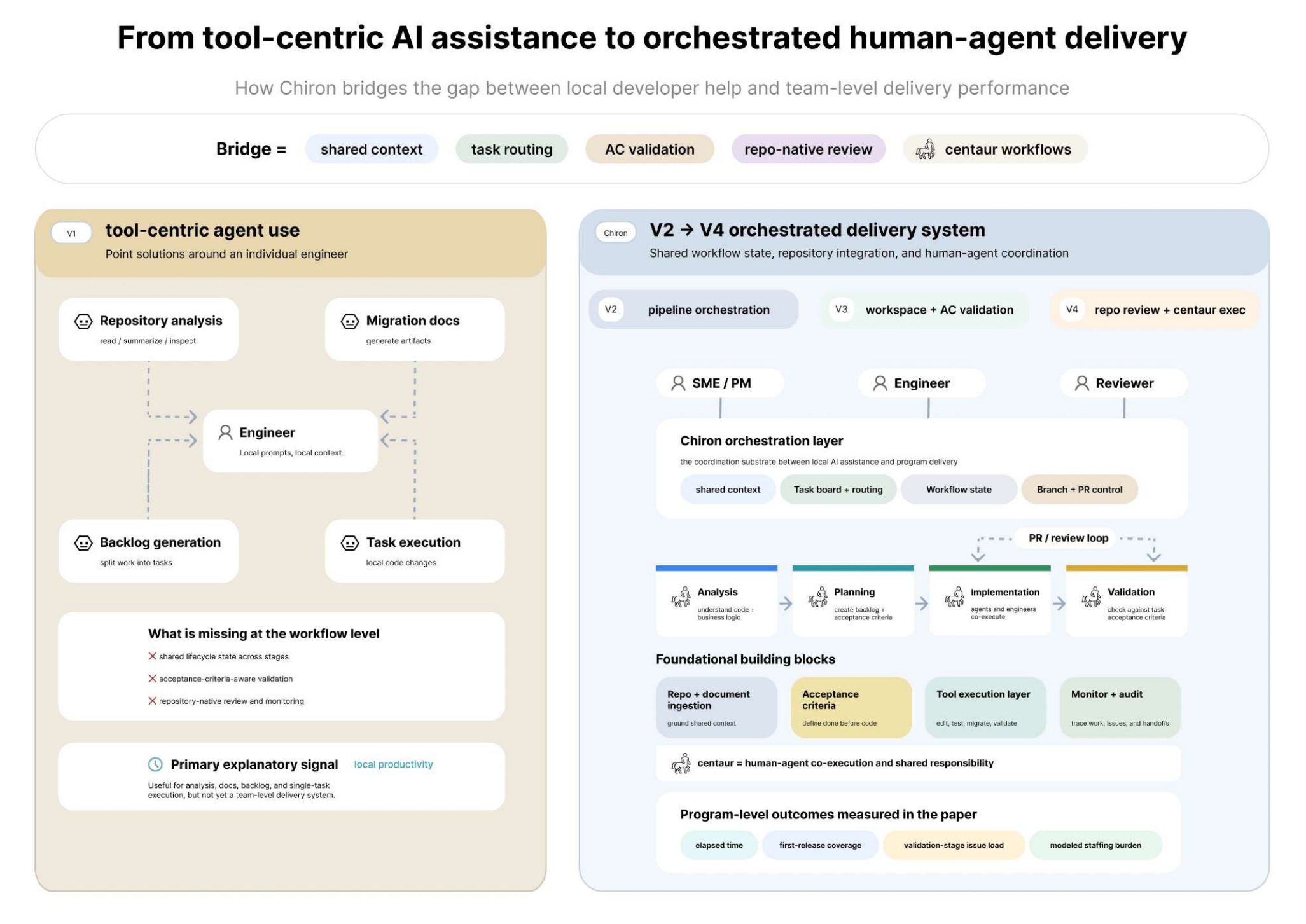}
\caption{Orchestrated agentic software delivery across analysis, planning, implementation, and validation stages, showing hybrid human--AI coordination.}
\label{fig:orchestrated-delivery}
\end{figure}

\section{Introduction}

Large language models have clearly changed the economics of \emph{local} programming work. Controlled and field studies have reported faster completion of programming tasks, measurable changes in throughput, and heterogeneous gains across experience levels and work settings \cite{peng2023copilot,ziegler2024copilot,cui2026highskilled}. At the same time, the measurement literature has repeatedly cautioned that software productivity is inherently multidimensional and should not be collapsed into a single proxy \cite{forsgren2021space}.

That tension matters for AI evaluation. A modern software program is not delivered by isolated code generation alone. It moves through technical understanding, planning, decomposition, implementation, verification, review, and release coordination. Yet much of the present empirical evidence still focuses either on individual task assistance \cite{peng2023copilot,ziegler2024copilot,cui2026highskilled,becker2025experienced} or on repository-level issue resolution benchmarks such as SWE-bench and SWE-agent \cite{jimenez2024swebench,yang2024sweagent,wang2024openhands}. Those evaluations are important, but they do not by themselves answer the managerial and scientific question of interest here: \emph{what happens when agents are embedded in an end-to-end delivery workflow used by a hybrid human-AI team?}

This paper examines that question in an industrial setting. We study Chiron, a team-level software delivery platform that evolved from tool-centric agent use (V1) toward increasingly integrated orchestration across documentation, backlog generation, acceptance-criteria validation, repository-native review, and hybrid execution. The dataset covers three real modernization programs and five delivery configurations. Unlike a controlled trial, the study does not isolate a single component or model; instead, it characterizes the behavior of an evolving delivery system deployed in practice.

The literature gives good reason to study orchestration directly. Project-level open-source evidence suggests that AI assistance can improve output while also increasing integration time, highlighting that team-level coordination costs need not move in lockstep with individual productivity gains \cite{song2024oss}. Repository-level agent research likewise points toward the importance of scaffolding, interfaces, context management, and verification loops rather than raw model capability alone \cite{jimenez2024swebench,yang2024sweagent,li2025deepcode,xu2025everything,jiang2025adaptation}. Long-horizon agentic systems outside software engineering exhibit a similar pattern: decomposition and local error correction become increasingly important as workflow length grows \cite{meyerson2025maker}.

The contribution of this paper is therefore descriptive and systems-oriented rather than causal. Specifically, we:
\begin{enumerate}[leftmargin=1.4em,itemsep=0.15em]
    \item define a measurement framework that cleanly separates \emph{observed delivery outcomes} from \emph{modeled staffing scenarios};
    \item report a retrospective longitudinal field study covering three real modernization programs and five delivery configurations; and
    \item show that the largest improvements appear only after orchestration, acceptance-criteria validation, and repository-native review are added to the delivery system.
\end{enumerate}

To prevent overclaiming, the paper is explicit about what the evidence does \emph{not} show. It does not provide a randomized causal estimate. It does not identify the contribution of any single foundation model, prompt, tool, or workflow feature. It does not measure post-release reliability or total lifecycle cost. What it does provide is a carefully reconstructed industrial record showing that progressively more orchestrated human-AI delivery configurations were associated with markedly shorter delivery times, lower downstream issue load, higher first-release coverage, and lower modeled staffing burden within one organization.

\section{Study Context and Research Questions}

\subsection{System under study}

Chiron is a team-level software delivery platform rather than a standalone coding assistant. The system supports repository ingestion, technology-stack and business-logic analysis, documentation synthesis, backlog generation, agent- and human-assigned execution, acceptance-criteria-based validation, pull-request generation, and repository-native review. The unit of interest in this paper is therefore not an isolated model completion. It is a delivery program that moves from technical understanding to validated first release.

\subsection{Platform evolution}

Table~\ref{tab:versions} summarizes the four agentic configurations benchmarked in the study. The versions should be read as successive delivery configurations, not as independent experimental treatments. The practical importance of this distinction is that later versions combine architectural maturation, workflow redesign, and organizational learning.

\begin{table}[H]
\caption{Versioned delivery configurations.}
\label{tab:versions}
\small
\begin{tabularx}{\textwidth}{@{}lYY@{}}
\toprule
Version & Dominant capabilities & Practical interpretation \\
\midrule
V1 & Tool-centric agent usage for repository analysis, migration documentation, backlog generation, and autonomous task execution, with limited integrated validation or repository-native review. & Earliest gains appear in analysis and planning, but downstream quality weakens. \\
V2 & CLI-orchestrated delivery pipelines for context preparation, implementation, and validation. & Architectural normalization yields only modest improvement over V1. \\
V3 & Shared web workspace, brainstorming interface, task board, automatic task pickup, and first-generation validation against task acceptance criteria. & Implementation and validation improve sharply once orchestration becomes task-centric. \\
V4 & Repository authentication, branch and PR workflows, code review, document ingestion, process monitoring, and hybrid human-agent execution. & Best observed balance of speed, first-release coverage, and downstream issue load. \\
\bottomrule
\end{tabularx}
\end{table}

The most conservative reading of V1 is that it approximates a \emph{tool-centric} mode of AI use: engineers interact with agents for local analysis, documentation, and task execution, but without a unified lifecycle workflow. V2 through V4 add progressively more coordination and repository integration. This makes the V1-to-V4 comparison particularly important because those versions share the same nominal staffing scenario while differing substantially in orchestration maturity.

\subsection{Benchmark programs}

The portfolio contains three modernization programs chosen because they differ in scale, technology mix, and legacy burden. Table~\ref{tab:projects} summarizes the programs.

\begin{table}[H]
\caption{Benchmark portfolio.}
\label{tab:projects}
\small
\begin{tabularx}{\textwidth}{@{}l l Y l@{}}
\toprule
Project & Approx. size & Modernization summary & Traditional baseline \\
\midrule
Bank application & $\sim$30k LOC & COBOL backend to Python, terminal frontend to Next.js, and database migration to PostgreSQL. & 10.0 weeks \\
ACAS accounting platform & $\sim$400k LOC & Large legacy accounting modernization. & 20.0 weeks \\
Mortgage application & $\sim$30k LOC & .NET 3 to .NET 8, Angular to React, and REST to GraphQL. & 6.0 weeks \\
\bottomrule
\end{tabularx}
\end{table}

These cases span both classic legacy migration and modern-stack modernization, reducing the risk that the observed pattern is unique to COBOL translation alone. They do \emph{not} constitute a representative sample of all software projects, and the external validity of the study should be interpreted accordingly.

\subsection{Research questions}

The analysis is organized around four descriptive research questions.

\begin{description}[leftmargin=1.4em,style=nextline]
\item[RQ1.] How do measured delivery durations change across platform evolution and relative to the traditional baseline?
\item[RQ2.] How do downstream validation outcomes and first-release coverage change across versions?
\item[RQ3.] Under explicit staffing scenarios, how large are the implied reductions in raw and senior-equivalent effort?
\item[RQ4.] Which parts of platform evolution are most plausibly associated with the largest observed gains?
\end{description}

\section{Study Design and Measurement}

\subsection{Design and unit of analysis}

This is a \emph{retrospective longitudinal field study}. The observational unit is a project-version cell $(p,v)$, where $p$ indexes one of the three modernization programs and $v \in \{\text{Traditional},\text{V1},\text{V2},\text{V3},\text{V4}\}$ indexes a delivery configuration. The complete dataset therefore comprises fifteen project-version cells.

The versions are not independent replications. They represent successive delivery configurations of the same platform family. For that reason, we report descriptive effect sizes and sensitivity analyses rather than null-hypothesis significance tests.

\subsection{Data sources and provenance}

The benchmark was compiled retrospectively from engineering records and practitioner recall. The source manuscript did not contain a cell-level provenance matrix distinguishing which values were recovered entirely from durable records and which required reconstruction from memory. We therefore adopt the most conservative interpretation of the evidence: the paper should be read as a retrospective industrial field study, not as an instrumented benchmark with uniformly contemporaneous telemetry.

The available source material supports project-version measurements for:
\begin{itemize}[leftmargin=1.4em,itemsep=0.15em]
    \item stage durations for analysis, planning, implementation, and validation;
    \item backlog task counts;
    \item issues reaching downstream validation;
    \item first-release requirement coverage; and
    \item team composition assumptions used for staffing-normalized effort scenarios.
\end{itemize}

For V4 only, the benchmark also records approximate counts of issues observed before downstream validation, allowing a limited analysis of review-stage containment.

\subsection{Delivery configurations and staffing scenarios}

To preserve cross-version comparability, the study measures four stages:
\[
S = \{\text{Analysis},\text{Planning},\text{Implementation},\text{Validation}\}.
\]

For project $p$, version $v$, and stage $s$, let $\tau_{p,v,s}$ denote measured duration in weeks. Total measured duration is

\[
T_{p,v} = \sum_{s \in S} \tau_{p,v,s}.
\]

Deployment is excluded from the quantitative core because that stage was not represented consistently across all versions. Its effects may still appear indirectly through validation duration, downstream issue load, and first-release coverage.

The paper uses two staffing scenarios. The traditional scenario fixes staffing at six people:
\[
\text{Traditional team} = \{1 \text{ architect}, 2 \text{ backend}, 1 \text{ frontend}, 2 \text{ QA}\}.
\]

The agentic scenario fixes staffing at five people:
\[
\text{Agentic team} = \{1 \text{ senior architect}, 2 \text{ junior AI operators}, 2 \text{ junior QA}\}.
\]

These scenarios are used only for modeled effort calculations. They are not direct labor-hour logs.

\subsection{Observed outcome measures}

The primary observed variables are:
\begin{itemize}[leftmargin=1.4em,itemsep=0.15em]
    \item total measured duration $T_{p,v}$;
    \item task count $N_{p,v}$;
    \item validation issues $I_{p,v}$, defined as issues that reached downstream validation; and
    \item first-release coverage $C_{p,v} \in [0,1]$, defined as the fraction of first-release requirements recorded as completed and accepted at initial handoff.
\end{itemize}

Because agentic configurations create more granular backlogs than the traditional baseline, raw issue counts are not directly comparable across versions. We therefore use a workload-normalized downstream quality measure:

\[
L_{p,v} = 100 \times \frac{I_{p,v}}{N_{p,v}},
\]

reported as \emph{validation-stage issue load per 100 tasks}. This metric should be interpreted as a downstream issue-escape measure, not as an intrinsic defect density. It reflects both defect creation and defect containment before validation.

For portfolio coverage, we use task-weighted aggregation:
\[
\bar{C}_v = \frac{\sum_{p \in P} C_{p,v} N_{p,v}}{\sum_{p \in P} N_{p,v}},
\]
where $P$ is the set of projects. Weighting by task volume avoids treating the 400k-LOC accounting modernization and the 30k-LOC mortgage modernization as equal-sized workloads.

\subsection{Modeled effort measures}

We report two staffing-normalized effort scenarios.

\paragraph{Raw person-days.} Let $h_v$ denote the assumed team headcount. With a five-day work week,
\[
E^{\mathrm{raw}}_{p,v} = 5 h_v T_{p,v},
\]
where $h_{\text{Traditional}} = 6$ and $h_{\text{V1--V4}} = 5$.

\paragraph{Senior-equivalent effort.} Let $\sigma_v$ denote senior-equivalent staffing. Under the benchmark's baseline weighting, each junior team member counts as $0.5$ senior equivalents, so the agentic team is modeled as $3$ SEE:
\[
E^{\mathrm{SEE}}_{p,v} = 5 \sigma_v T_{p,v},
\]
where $\sigma_{\text{Traditional}} = 6$ and $\sigma_{\text{V1--V4}} = 3$.

These modeled effort values are useful for staffing scenarios and executive planning, but they are not direct empirical observations. To make that dependence explicit, Section~\ref{sec:sensitivity} reports sensitivity to alternative junior-to-senior weights.

\subsection{Analysis strategy}

The analysis is descriptive. We report project-level values, portfolio aggregates, and simple ratios or percentage changes. No significance tests are used because the dataset is small, retrospective, and serially dependent across versions. Robustness is assessed in three ways:
\begin{enumerate}[leftmargin=1.4em,itemsep=0.15em]
    \item by showing project-level trajectories rather than only portfolio totals;
    \item by comparing V4 not only to the traditional baseline but also to V1, which reflects individual human use of agents without orchestration; and
    \item by conducting sensitivity analyses for senior-equivalent weighting and leave-one-project-out portfolio aggregation.
\end{enumerate}

\section{Results}

\subsection{Portfolio evolution across delivery configurations}

Table~\ref{tab:portfolio} summarizes the portfolio across all five delivery configurations. The ``weeks'' column is the \emph{sum of project durations} rather than a literal concurrent portfolio makespan.

\begin{table}[H]
\caption{Portfolio evolution across delivery configurations. Coverage values are recalculated directly from the project tables and task weights.}
\label{tab:portfolio}
\small
\begin{tabularx}{\textwidth}{@{}lrrrrr@{}}
\toprule
Version & Weeks & Raw person-days & SEE-days & Issue load / 100 tasks & Task-weighted coverage \\
\midrule
Traditional & 36.0 & 1080.0 & 1080.0 & 8.03 & 77.0\% \\
V1 & 28.6 & 715.0 & 429.0 & 8.63 & 52.6\% \\
V2 & 26.4 & 660.0 & 396.0 & 8.17 & 57.6\% \\
V3 & 13.8 & 345.0 & 207.0 & 4.19 & 83.4\% \\
V4 & 9.3 & 232.5 & 139.5 & 2.09 & 90.5\% \\
\bottomrule
\end{tabularx}
\end{table}

Relative to the traditional baseline, V4 reduces summed project duration from 36.0 to 9.3 weeks, a 3.87$\times$ speedup and a 74.2\% reduction in measured elapsed time. Under the benchmark's staffing scenarios, modeled raw effort falls by 78.5\% and modeled senior-equivalent effort falls by 87.1\%. At the same time, validation-stage issue load falls from 8.03 to 2.09 issues per 100 tasks, and task-weighted first-release coverage rises from 77.0\% to 90.5\%.

The full trajectory is as important as the endpoint. V1 and V2 improve speed early, especially in analysis and planning, but they worsen downstream issue load and coverage relative to the traditional baseline. V3 and V4 reverse that pattern. By V3, the portfolio is already faster than the baseline and better on both normalized downstream quality and coverage; V4 extends that improvement further.

\begin{table}[H]
\caption{Two informative comparisons for the mature V4 configuration. The V1-to-V4 contrast is especially useful because both configurations share the same nominal agentic staffing scenario.}
\label{tab:summary}
\small
\begin{tabularx}{\textwidth}{@{}lrrrrrr@{}}
\toprule
Contrast & Speedup & Time red. & Raw-effort red. & SEE red. & Issue-load red. & Coverage change \\
\midrule
Traditional $\rightarrow$ V4 & 3.87$\times$ & 74.2\% & 78.5\% & 87.1\% & 74.0\% & +13.4 pp \\
V1 $\rightarrow$ V4 & 3.08$\times$ & 67.5\% & 67.5\% & 67.5\% & 75.8\% & +37.9 pp \\
\bottomrule
\end{tabularx}
\end{table}

The V1-to-V4 comparison strengthens the orchestration interpretation. Because V1 through V4 use the same nominal five-person, three-SEE staffing scenario, the V1-to-V4 reduction is not driven by a staffing-composition change. From V1 to V4, the portfolio becomes 3.08$\times$ faster, validation-stage issue load falls by 75.8\%, and task-weighted first-release coverage rises by 37.9 percentage points. In other words, later gains do not merely reflect ``having agents''; they appear only once the agents are embedded in a more structured delivery workflow.

\subsection{Time compression}

\begin{figure}[H]
    \centering
    \includegraphics[width=0.93\textwidth]{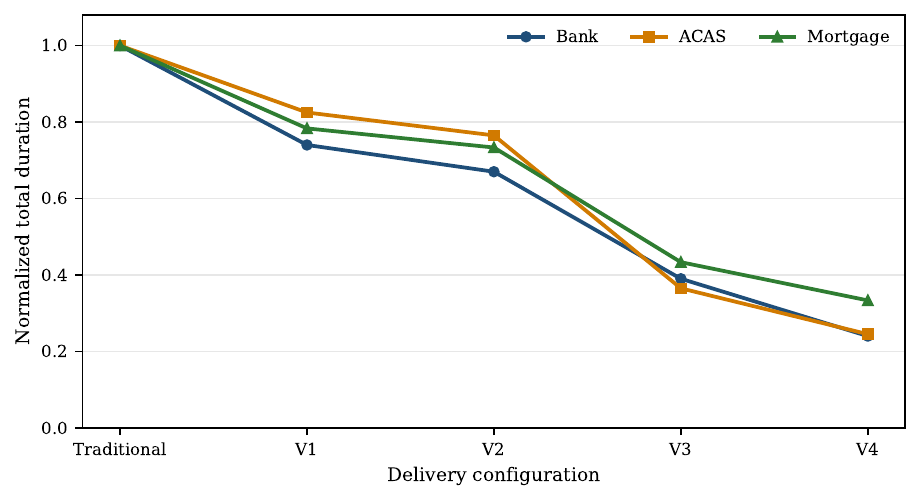}
    \caption{Total measured duration normalized to each project's traditional baseline. Values below 1 indicate faster delivery.}
    \label{fig:duration}
\end{figure}

Figure~\ref{fig:duration} shows the project-level duration trajectory normalized to each project's traditional baseline. Every project follows the same high-level pattern. The earliest improvements arrive in V1 and V2, and the sharpest additional gains arrive in V3 and V4.

This pattern matters substantively. If the principal effect of AI were merely faster local code generation, one might expect V1 already to dominate the later versions. It does not. The data instead suggest a two-step curve: first, understanding and planning compress; later, the system becomes stable enough for faster implementation to \emph{coexist} with lower downstream issue load.

\subsection{Quality and first-release outcomes}

\begin{figure}[H]
    \centering
    \includegraphics[width=0.98\textwidth]{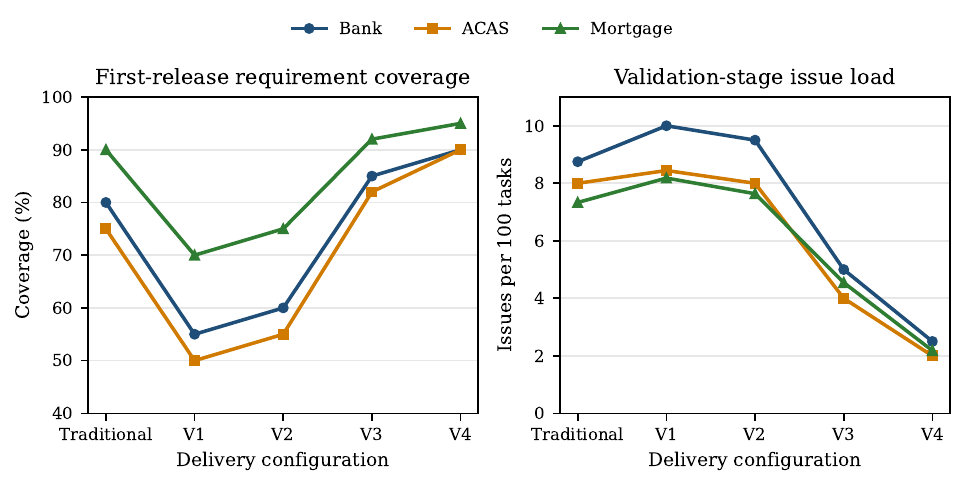}
    \caption{First-release coverage and validation-stage issue load across delivery configurations. The issue metric should be interpreted as a workload-normalized downstream escape measure rather than intrinsic defect density.}
    \label{fig:quality}
\end{figure}

Figure~\ref{fig:quality} reports the two observed quality-adjacent outcomes used in the study: first-release requirement coverage and validation-stage issue load. The early versions are best described as speed-first systems. They reduce stage durations before the workflow has adequate downstream controls.

By V4, every project exceeds its traditional baseline on both measures. Bank improves from 8.75 to 2.50 issues per 100 tasks while coverage rises from 80\% to 90\%. ACAS moves from 8.00 to 2.00 while coverage rises from 75\% to 90\%. Mortgage moves from 7.33 to 2.18 while coverage rises from 90\% to 95\%.

These improvements should nonetheless be interpreted carefully. The quality metric counts issues that \emph{reach validation}. A lower value can arise because fewer issues are created, because more are caught earlier, or both. Section~\ref{sec:containment} shows direct evidence that review-stage containment increases in V4.

\subsection{Stage-specific behavior}

\begin{figure}[H]
    \centering
    \includegraphics[width=0.90\textwidth]{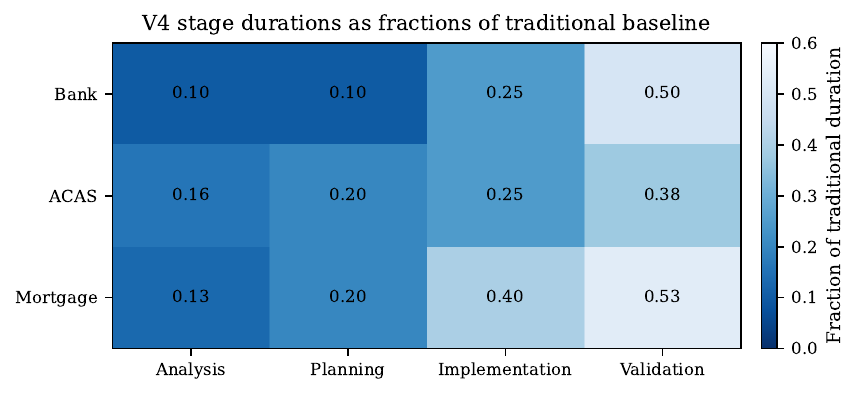}
    \caption{V4 stage durations as fractions of the traditional baseline. Lower values indicate shorter duration in the measured stage.}
    \label{fig:stage}
\end{figure}

Figure~\ref{fig:stage} compares V4 against the traditional baseline at the stage level. Each cell is the ratio $\tau_{p,\text{V4},s} / \tau_{p,\text{Traditional},s}$. Lower values indicate stronger compression of the measured stage.

The largest and most consistent reductions occur in analysis and planning. For the Bank program, analysis drops from 2.0 weeks to 0.2 and planning drops from 2.0 weeks to 0.2. ACAS falls from 5.0 to 0.8 weeks in analysis and from 3.0 to 0.6 in planning. Mortgage falls from 1.5 to 0.2 weeks in analysis and from 1.0 to 0.2 in planning. The interpretation is that the system reduces time-to-engineering-readiness, not only time-to-code.

Implementation and validation also shorten, but less uniformly. That difference is plausible: later stages remain closer to the codebase, test harness, and defect surface, where reliability limits tend to bind harder than in analysis and planning.

\subsection{Defect containment through review}
\label{sec:containment}

The V4 configuration introduces repository-native review and PR workflows, making it possible to estimate how many issues are caught before downstream validation. The benchmark records approximate pre-validation issue counts of 40 for Bank, 188 for ACAS, and 23 for Mortgage, of which 20, 90, and 12 respectively reach downstream validation. That implies review-stage containment rates of 50.0\%, 52.1\%, and 47.8\%, with a portfolio-weighted containment rate of approximately 51.4\%.

This finding is important for interpretation. The main downstream quality metric is not a pure measure of defect generation; it is partly a measure of \emph{where} defects are detected. V4 improves delivery not only by producing better outputs, but also by moving part of defect discovery from late-stage validation to earlier review, where rework is usually cheaper.

\subsection{Sensitivity of modeled senior-equivalent effort}
\label{sec:sensitivity}

Because senior-equivalent effort depends on the assumed junior-to-senior weighting, Table~\ref{tab:sensitivity} reports a simple sensitivity analysis. Let each junior team member count as $w$ senior equivalents, with the agentic team therefore modeled as $1 + 4w$ SEE.

\begin{table}[H]
\caption{Sensitivity of modeled V4 senior-equivalent effort to alternative junior weights. The traditional baseline remains fixed at 6 SEE.}
\label{tab:sensitivity}
\small
\begin{tabularx}{\textwidth}{@{}rrrr@{}}
\toprule
Junior weight $w$ & Agentic SEE staffing & V4 SEE-days & Reduction vs. traditional \\
\midrule
0.25 & 2.0 & 93.0 & 91.4\% \\
0.50 & 3.0 & 139.5 & 87.1\% \\
0.75 & 4.0 & 186.0 & 82.8\% \\
1.00 & 5.0 & 232.5 & 78.5\% \\
\bottomrule
\end{tabularx}
\end{table}

The modeled reduction is large across all reasonable weights. Even if each junior team member is counted as a full senior equivalent ($w = 1.00$), V4 still reduces portfolio SEE-days by 78.5\% relative to the traditional baseline because elapsed time falls so sharply. This does not eliminate the need for better utilization data, but it shows that the staffing conclusion is not uniquely dependent on the baseline $w = 0.50$ assumption.

\section{Discussion}

\subsection{What the evidence supports}

The strongest claim supported by the study is descriptive: within one organization, across three modernization programs, progressively more orchestrated human-AI delivery configurations were associated with much shorter delivery times, lower downstream validation-stage issue load, higher first-release coverage, and lower modeled staffing burden.

That claim is substantively meaningful. The effect sizes are large, they point in the same direction across all three projects, and the later versions improve both speed and downstream outcomes rather than merely trading one for the other. They also align with a systems-level view of software delivery: value is created not only by code generation, but by structuring information flow, decomposing work, validating against explicit acceptance criteria, and reviewing changes before errors cascade \cite{li2025deepcode,xu2025everything,jiang2025adaptation,meyerson2025maker}.

At the same time, the evidence does \emph{not} justify a stronger causal claim such as ``orchestration alone causes the gains.'' Across V1 through V4, multiple things can change at once: model capability, prompts, context-construction procedures, workflow maturity, review habits, and operator skill. The correct interpretation is therefore system-level rather than component-level.

\subsection{Why the V1-to-V4 contrast matters}

Among the comparisons in the paper, the V1-to-V4 contrast is the least confounded by staffing assumptions because both configurations share the same nominal five-person, three-SEE staffing scenario. Under that comparison, time and modeled effort both fall by 67.5\%, while downstream issue load falls by 75.8\% and coverage rises by 37.9 percentage points.

This matters because it distinguishes \emph{agent availability} from \emph{agent orchestration}. V1 already includes local agent use for analysis, documentation, backlog creation, and autonomous task execution. Yet V1 is not the best configuration. The strongest results arrive only after the workflow becomes acceptance-criteria-aware, repository-native, and review-aware. That pattern is more consistent with an orchestration thesis than with a simple ``better autocomplete'' thesis.

\subsection{Implications for AI evaluation in software engineering}

The study also highlights a methodological point for the broader literature. Much current evaluation still focuses on repository-level issue solving, coding benchmarks, or developer-local assistance \cite{jimenez2024swebench,yang2024sweagent,wang2024openhands}. Those evaluations remain essential, but they do not fully characterize organizational delivery performance. The empirical literature already suggests that project-level coordination can attenuate, amplify, or even reverse local productivity gains depending on the work context \cite{song2024oss,becker2025experienced}.

For industrial deployment, that implies at least three measurement rules.

First, observed and modeled metrics should be separated. Elapsed time, issue counts, and coverage are empirical outcomes; person-days and senior-equivalent days are scenario-based interpretations of those outcomes.

Second, downstream quality should be staged. A mature delivery evaluation should report defects found in review, in validation, and after release rather than collapsing them into a single count.

Third, denominators should be chosen carefully. When an intervention changes task granularity, metrics normalized ``per task'' become useful but imperfect. The paper therefore treats validation-stage issue load as a pragmatic normalization rather than a definitive quality measure.

\section{Threats to Validity and Scope Conditions}

The principal threats to validity are structural rather than cosmetic, and they should be taken seriously.

\paragraph{Retrospective reconstruction and provenance heterogeneity.} The benchmark was compiled retrospectively from engineering records and practitioner recall. Because the available source material does not provide a cell-by-cell provenance table, some measurements may be more durable than others. This risk mainly affects interpretability and uncertainty, not the arithmetic of the reported tables.

\paragraph{Single-organization setting and project selection.} All three programs come from one organization and one platform family. The results may therefore reflect organization-specific engineering practices, review culture, or project mix. External validity is limited.

\paragraph{Longitudinal platform evolution.} The versions are successive delivery configurations, not isolated ablations. Improvements may reflect cumulative workflow learning, improved integration, or changes in underlying models and prompts. The study is appropriate for evaluating platform evolution, but not for component-level attribution.

\paragraph{Task-granularity denominator drift.} Agentic configurations produce more granular backlogs than the traditional baseline. The issue-load metric addresses this partially by normalizing per 100 tasks, but a task in one configuration is not necessarily semantically equivalent to a task in another.

\paragraph{Observed downstream issues are not total defects.} The main quality metric counts issues reaching validation. V4's repository-native review likely diverts some issues out of that channel, which is valuable operationally but means the metric captures defect escape to validation rather than total defect generation.

\paragraph{Coverage is a release-readiness metric.} First-release coverage captures what ships at initial handoff. It does not measure post-release convergence, maintainability, or total lifecycle cost.

\paragraph{Modeled effort is scenario-dependent.} Raw person-days and SEE-days are derived from staffing assumptions rather than direct labor-hour logs. The sensitivity analysis reduces, but does not eliminate, that limitation.

\section{Conflict of Interest and Author Positionality}

Both authors hold executive roles at Taller Technologies, the company that developed Chiron. This creates clear incentives toward favorable interpretation. To mitigate that risk, the paper reports all project-version tables available in the source material, separates observed from modeled metrics, avoids causal language, and includes sensitivity analyses. No independent external audit or blinded adjudication was performed in the present study. Readers should interpret the paper as an industrial field report with explicit conflict-of-interest disclosure.

\section{Conclusion}

This paper reported a retrospective longitudinal field study of an industrial human-AI software delivery platform across three real modernization programs and five delivery configurations. The core result is not merely that AI can accelerate local coding tasks. Rather, the data suggest that the largest gains appear when agents are embedded in a structured delivery workflow that combines decomposition, acceptance-criteria validation, repository-native review, and hybrid human-agent execution.

Within the observed portfolio, the mature V4 configuration is associated with shorter measured delivery time, lower downstream validation-stage issue load, higher first-release coverage, and lower modeled staffing burden than both the traditional baseline and the early tool-centric agentic configuration V1. Those effect sizes are large enough to be practically important even under a conservative descriptive interpretation.

The paper's limitations are real: retrospective provenance, single-organization scope, changing task granularity, and the absence of post-release defect data all constrain what can be claimed. But those limitations do not erase the central empirical signal. If replicated with stronger instrumentation and broader project coverage, the findings would support a shift in how the field evaluates AI for software engineering: from isolated code-generation events toward the design and measurement of orchestrated delivery systems.

\appendix

\section{Complete Benchmark Tables}

\begin{table}[H]
\caption{Bank application benchmark across delivery configurations.}
\small
\begin{tabularx}{\textwidth}{@{}lrrrrrrrr@{}}
\toprule
Version & Analysis & Planning & Impl. & Val. & Total & Tasks & Issues & Coverage \\
\midrule
Traditional & 2.0 & 2.0 & 4.0 & 2.0 & 10.0 & 400 & 35 & 80\% \\
V1 & 1.0 & 0.4 & 3.0 & 3.0 & 7.4 & 800 & 80 & 55\% \\
V2 & 0.8 & 0.4 & 2.5 & 3.0 & 6.7 & 800 & 76 & 60\% \\
V3 & 0.6 & 0.3 & 1.5 & 1.5 & 3.9 & 800 & 40 & 85\% \\
V4 & 0.2 & 0.2 & 1.0 & 1.0 & 2.4 & 800 & 20 & 90\% \\
\bottomrule
\end{tabularx}
\end{table}

\begin{table}[H]
\caption{ACAS accounting benchmark across delivery configurations.}
\small
\begin{tabularx}{\textwidth}{@{}lrrrrrrrr@{}}
\toprule
Version & Analysis & Planning & Impl. & Val. & Total & Tasks & Issues & Coverage \\
\midrule
Traditional & 5.0 & 3.0 & 8.0 & 4.0 & 20.0 & 2500 & 200 & 75\% \\
V1 & 2.5 & 1.0 & 6.0 & 7.0 & 16.5 & 4500 & 380 & 50\% \\
V2 & 2.2 & 0.9 & 5.2 & 7.0 & 15.3 & 4500 & 360 & 55\% \\
V3 & 1.5 & 0.6 & 3.2 & 2.0 & 7.3 & 4500 & 180 & 82\% \\
V4 & 0.8 & 0.6 & 2.0 & 1.5 & 4.9 & 4500 & 90 & 90\% \\
\bottomrule
\end{tabularx}
\end{table}

\begin{table}[H]
\caption{Mortgage application benchmark across delivery configurations.}
\small
\begin{tabularx}{\textwidth}{@{}lrrrrrrrr@{}}
\toprule
Version & Analysis & Planning & Impl. & Val. & Total & Tasks & Issues & Coverage \\
\midrule
Traditional & 1.5 & 1.0 & 2.0 & 1.5 & 6.0 & 300 & 22 & 90\% \\
V1 & 0.8 & 0.4 & 1.5 & 2.0 & 4.7 & 550 & 45 & 70\% \\
V2 & 0.7 & 0.4 & 1.3 & 2.0 & 4.4 & 550 & 42 & 75\% \\
V3 & 0.4 & 0.2 & 1.0 & 1.0 & 2.6 & 550 & 25 & 92\% \\
V4 & 0.2 & 0.2 & 0.8 & 0.8 & 2.0 & 550 & 12 & 95\% \\
\bottomrule
\end{tabularx}
\end{table}

\begin{table}[H]
\caption{Modeled effort and observed validation-stage issue load by project and version.}
\small
\begin{tabularx}{\textwidth}{@{}llrrr@{}}
\toprule
Project & Version & Raw person-days & SEE-days & Issue load / 100 tasks \\
\midrule
Bank & Traditional & 300.0 & 300.0 & 8.75 \\
Bank & V1 & 185.0 & 111.0 & 10.00 \\
Bank & V2 & 167.5 & 100.5 & 9.50 \\
Bank & V3 & 97.5 & 58.5 & 5.00 \\
Bank & V4 & 60.0 & 36.0 & 2.50 \\
ACAS & Traditional & 600.0 & 600.0 & 8.00 \\
ACAS & V1 & 412.5 & 247.5 & 8.44 \\
ACAS & V2 & 382.5 & 229.5 & 8.00 \\
ACAS & V3 & 182.5 & 109.5 & 4.00 \\
ACAS & V4 & 122.5 & 73.5 & 2.00 \\
Mortgage & Traditional & 180.0 & 180.0 & 7.33 \\
Mortgage & V1 & 117.5 & 70.5 & 8.18 \\
Mortgage & V2 & 110.0 & 66.0 & 7.64 \\
Mortgage & V3 & 65.0 & 39.0 & 4.55 \\
Mortgage & V4 & 50.0 & 30.0 & 2.18 \\
\bottomrule
\end{tabularx}
\end{table}

\section{Additional Robustness and Interpretation Tables}

{\small
\setlength{\LTleft}{0pt}
\setlength{\LTright}{0pt}
\begin{longtable}{@{}p{0.17\textwidth}p{0.12\textwidth}p{0.31\textwidth}p{0.32\textwidth}@{}}
\caption{Metric taxonomy used in the paper.}\\
\toprule
Metric & Type & Definition & Interpretation caveat \\
\midrule
\endfirsthead
\toprule
Metric & Type & Definition & Interpretation caveat \\
\midrule
\endhead
Total duration $T_{p,v}$ & Observed & Sum of measured analysis, planning, implementation, and validation durations. & Excludes deployment because that stage was not represented consistently across versions. \\
Task count $N_{p,v}$ & Observed & Number of recorded work items in the project backlog for the project-version cell. & Task granularity changes materially across delivery configurations. \\
Validation issues $I_{p,v}$ & Observed & Issues reaching downstream validation. & Captures defect escape to validation, not total defects created anywhere in the pipeline. \\
Validation-stage issue load $L_{p,v}$ & Derived from observed measures & $100 \times I_{p,v}/N_{p,v}$, reported as issues per 100 tasks. & A workload-normalized escape metric; not equivalent to intrinsic defect density because the denominator is intervention-sensitive. \\
First-release coverage $C_{p,v}$ & Observed & Fraction of first-release requirements recorded as completed and accepted at initial handoff. & Sensitive to requirement scoping and adjudication practices. \\
Raw person-days $E^{\mathrm{raw}}_{p,v}$ & Modeled & $5 h_v T_{p,v}$ under full-time staffing with $h_{\mathrm{Traditional}} = 6$ and $h_{\mathrm{Agentic}} = 5$. & A scenario-based effort estimate, not directly logged labor-hours. \\
SEE-days $E^{\mathrm{SEE}}_{p,v}$ & Modeled & $5 \sigma_v T_{p,v}$ with $\sigma_{\mathrm{Traditional}} = 6$ and $\sigma_{\mathrm{Agentic}} = 3$ under the benchmark's baseline weighting. & Useful for staffing scenarios, but dependent on the chosen junior-to-senior weighting. \\
\bottomrule
\end{longtable}
}

\begin{table}[H]
\caption{Leave-one-project-out sensitivity for the V4 versus traditional portfolio comparison. Duration ratio is V4 divided by traditional, and issue-load ratio is the same ratio for validation-stage issue load.}
\small
\begin{tabularx}{\textwidth}{@{}lrrr@{}}
\toprule
Subset & Duration ratio & Issue-load ratio & Coverage change \\
\midrule
All projects & 0.258 & 0.260 & +13.4 pp \\
Exclude Bank & 0.265 & 0.255 & +13.9 pp \\
Exclude ACAS & 0.275 & 0.291 & +7.8 pp \\
Exclude Mortgage & 0.243 & 0.256 & +14.3 pp \\
\bottomrule
\end{tabularx}
\end{table}

\begin{table}[H]
\caption{Approximate V4 review containment. Pre-review issues are the benchmark's recorded issues observed before downstream validation.}
\small
\begin{tabularx}{\textwidth}{@{}lrrr@{}}
\toprule
Project & Pre-review issues & Validation issues & Containment rate \\
\midrule
Bank & 40 & 20 & 50.0\% \\
ACAS & 188 & 90 & 52.1\% \\
Mortgage & 23 & 12 & 47.8\% \\
\bottomrule
\end{tabularx}
\end{table}

\end{document}